      [1999/12/01 v1.4c Il Nuovo Cimento]
\documentclass[varenna]{cimento}
\usepackage{graphicx}  
\usepackage{epsf}
\title{SZ cluster science with the Planck HFI experiment}
\author{Steen H. Hansen}

\institute{Physich Institute, Winterthurerstrasse 190, 8057 Zurich, Switzerland}
\begin{document}
\newcommand{\bi}{\bibitem}

\maketitle

\begin{abstract}
In the near future the Planck satellite~\cite{planck} will gather
impressive information about the anisotropies of the cosmic microwave
background and about the galaxy clusters which perturb that signal. We
will here review the ability of Planck to extract information about
these galaxy clusters through the Sunyaev-Zeldovich imprint. We will
conclude that Planck will provide a catalogue of galaxy clusters, which
will be very useful for future targeted observations. We will explain
why Planck will not be very good in extracting detailed information
about individual clusters, except for the dominating Compton
parameter, $y$, which will be measured to a few percent for individual
clusters. In this last point I am being rather conservative, but that
will leave space for pleasant surprises when the Planck data will be
analysed.
\end{abstract}

\section{Introduction}
The Planck satellite will be a real piece of art and an amazing piece
of hardware. It will provide us with information about the cosmic
background radiation (CMB) beyond the wildest imagination of most
people, surely beyond mine.  The sensitive detector will be sent far
into space, over a million kilometers from earth, where it will be
better protected from the Sun and Earth radiation. The satellite will
be spinning and carefully mapping the whole sky. Thus it will give us
accurate data about the full CMB sky, and about all the obstructions
and contaminations between us and the last scattering surface of the
CMB.

Many research years have been, and will still be, devoted to the
analysis and interpretation (and possibly even understanding) of the
CMB signal, and there is no question that the main scientific goal of
Planck will be to carefully study the CMB. In that sense all the
foregrounds, such as galaxy clusters and dust, are just contaminations
which will have to be carefully removed. I am going to be discussing
if we can use this removed ``contamination'' for anything useful.  It
is rather amusing that such beautiful structures as galaxies and
galaxy clusters are just considered to be contamination by many CMB
people.

The analysis of the Planck CMB data will determine many of the
cosmological parameters very accurately. This includes dark matter and
dark energy parameters as described by R. Bean and A. Melchiorri in
these proceedings. Numerous papers have reviewed the expected
precision with which Planck will determine such parameters, including
refs.~\cite{balbi,melch,bowen}.

One important strength of Planck is that it will have 9 observing
frequencies, and as we will see below this will allow a separation of the
CMB signal from the contaminations.
The actual construction and optimization of performance
of the Planck detector is a very complicated process, and I refer to
the lectures by P. de Bernardis and J. M. Lamarre in these proceedings
for all the details.
There are also recent nice reviews on the two 
instruments on-board Planck~\cite{lfi,hfi}, 
see also~\cite{villa,Yurchenko,naselsky}. 

Planck will find of the order 10 thousand massive galaxy clusters out
to a redshift of $z \sim 1.5$.  This cluster catalogue will be
deeper and larger (all sky survey) 
than any other existing catalogue. This
cluster catalogue will be very useful on its own and for future follow-up
observations. It is the goal of this lecture to discuss 
at an elementary level
how this will be achieved, and secondly we will discuss
more carefully what Planck will {\em not} be able to do.
Several groups have addressed similar issues already, see 
e.g. refs.~\cite{aghanim97,white03,geisbuesch}.

\section{Extracting the CMB}
The first question we have to ask is naturally {\em how bad are the
contaminations?} Or in other words, will we be able to extract the CMB
signal accurately enough to do proper CMB parameter extraction?

\begin{figure}
\begin{center}
\epsfxsize=8.5cm
\epsfysize=6.5cm
\epsffile{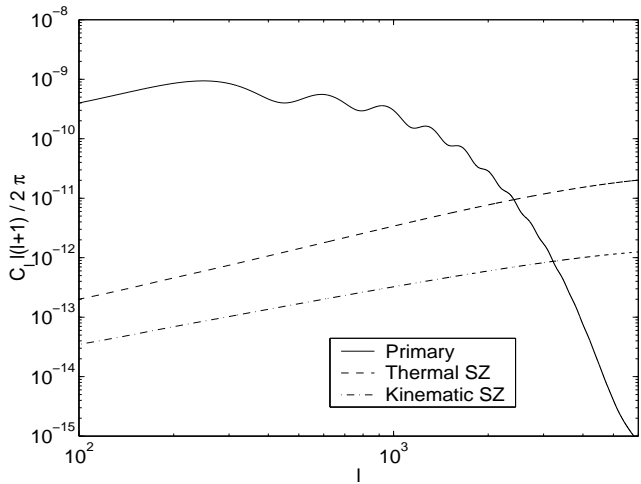}
\end{center}
\caption{The CMB power spectrum is seriously contaminated by the SZ
power for $l>2000$. The first question is, if this poses a serious
problem for CMB analysis? Figure taken from ref.\cite{rephaeli01}.}
\label{fig:rephaeli.power}
\end{figure}

There are many foreground components which will have to be removed,
these include infrared galaxies, dust, synchrotron radiation, free-free
emission, radio sources and the Sunyaev-Zeldovich effect. It turns
out that the dominating contamination at the CMB frequencies 
is the SZ effect, and
it is therefore crucial that we can separate the CMB signal from the
SZ signal. We will 
discuss the physics of the SZ effect in the next section. 
As is clear from fig.~\ref{fig:rephaeli.power} 
(from ref.~\cite{rephaeli01}, see also ref.~\cite{sadehrephaeli})
the power spectrum
of the CMB is dominated by the SZ effect for $l>2000$. This is 
very important for the CMB analyses of inflationary parameters, because
if we wish to allow e.g. a variation of the spectral index, then a
contamination free 
high-l part of the power spectrum is crucial~\cite{hannestad}.

Now, how is this component separation done?  The different signals
will have different frequency dependence.  For instance, the
unresolved radio sources will have most power at low frequencies
(30-70 GHz), and the dust will have most power at high
frequencies (545-857 GHz). Fortunately the SZ signal will be sitting
right in the middle, with most power on frequencies 143-353 GHz. Thus,
if we have several observing frequencies then we can use the different
frequency dependence of the signals to separate them.  
For an excellent review of the different contaminations see~\cite{bouchet}.
As mentioned in
the introduction, Planck will have 9 observing frequencies between
$30$ and $857$ GHz. In this way Planck can very carefully separate all
the different signals, hence removing the contaminations from the CMB
signal. The frequencies important for SZ science, 143-353 GHz 
are sitting on the HFI instrument~\cite{hfi}.

\begin{figure}
\begin{center}
\epsfxsize=8.5cm
\epsfysize=6.5cm
\epsffile{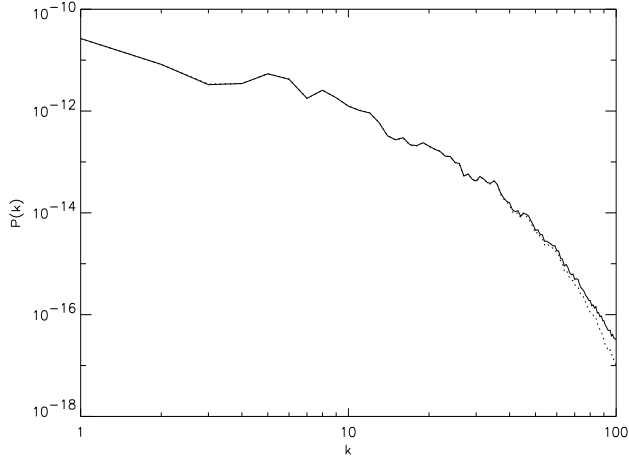}
\end{center}
\caption{The dotted line is the true CMB signal, and the full line is
the CMB signal plus non-removed SZ contamination.
The CMB power is well recovered from the contaminated signal.
Here the figure uses $k \approx 28\cdot l$, and the Power is the ``plane''
approximation of the normal $C_l$ (Fourier coefficients instead of spherical 
harmonics). Figure taken from ref.~\cite{diego}.}
\label{fig:diego.power}
\end{figure}

Already years ago it was understood that this component separation works
so well that the SZ signal will not be a major obstruction to CMB
science~\cite{heranz02,diego}, and this has been confirmed by several 
much more advanced studies~\cite{geisbuesch,heranz04}. An early figure showing
that the reconstructed CMB power is sufficiently accurate it presented
in fig.~\ref{fig:diego.power}

One point deserves a few comments, namely that the kinetic SZ effect has
the {\em same} spectral behaviour as the CMB signal itself. How can
one then separate these two? Basically one has to take advantage of
the fact that the thermal and kinetic SZ effects are spatially
correlated. Thus if you find a cluster (through the thermal effect)
then you can expect a kinetic effect to be right at that point in space 
- and at that point only.  You can therefore interpolate (guess) the
surrounding CMB signal through those points, and then you are left
with the kinetic effect~\cite{aghanimforni}.  
This method seems to work surprisingly well.

We may summarize this section by stating that even though the SZ
signal is very large, then it can be removed sufficiently accurately
that the CMB analysis can be performed with no problem. This was
the main goal of Planck, to do CMB analysis, so everyone is happy.
The next question to ask is whether we can use the removed SZ 
signal itself for a further cluster analysis?

\section{The Sunyaev-Zeldovich effect}
\label{sec:sz}
Let us make a small break from Planck and briefly discuss the SZ effect
itself.

Galaxy clusters typically have temperatures of the order keV, $T =
1-10$ keV. Therefore a CMB photon which traverses the cluster and
happens to Compton scatter off a hot electron will get increased
momentum. This up-scattering of CMB photons, which results in a small
change in the intensity of the cosmic microwave background, is known as
the Sunyaev-Zeldovich effect, and was predicted just over 30 years
ago~\cite{sz72}.  The first radiometric observations came few years
later~\cite{gull76,lake77}, and while recent years have seen an
impressive improvement in observational techniques and
sensitivity~\cite{laroque02,coma}, the near future observations
will see another boost in sensitivity by orders of magnitude.

From a theory point of view, the importance of a correct
treatment of the electron distribution function was emphasized about
25 years ago~\cite{wright79}, but an exact calculation of the SZ
effect was made only recently~\cite{dolgov00}.
Numerous groups have considered expansion in temperature, which works
accurately enough, and all the details are presented by
Y. Rephaeli and N. Itoh in these proceedings, and in the 
references~\cite{rephaeli95,shimon,ItohV,nozawa}.

In principle the relativistic corrections from a high electron
temperature can be measured with a multi-frequency observation, and
hence one can use {\em purely the SZ effect to determine the cluster
temperature}~\cite{pointec,hansen02}. 
This method is complementary to X-ray 
observations, because the SZ  temperature determination, which
simultaneously measures the electron density, $n_e$, does not depend on
an unknown clumping factor, as does the X-ray detection.
Furthermore, the SZ effect depends on very simple physics,
whereas understanding the X-ray signal is very complicated.

We have analysed existing SZ data and made the first cluster
temperature detections using purely the SZ
observations~\cite{hansen02,sasz}. 
These temperature detections have large
error-bars due to the limited sensitivity of present day observations,
but it has hereby been shown that this method of temperature
determination works, and this method can become a powerful tool for
cluster studies with future dedicated SZ surveys, and we will later
ask how well Planck will do here.

\section{Finding the clusters}
Most clusters have angular size of about 2 arcmin, so since Planck
channels will have angular resolution of 5 arcmin for 217 and 353 GHz
channels (and 7 arcmin for 143 GHz), then the individual clusters will
not be spatially resolved (except a few very nearby
ones)~\cite{white03}.  Whatever characteristics of the cluster we are
going to measure will be averaged over the full cluster.  At some
stage one will have to ask {\em how} this average is done, is it an
average over mass, an average over X-ray intensity, or an average over
SZ-intensity? Naturally there will be a difference between averages
over X-ray or SZ intensity, and one should therefore be rather careful
with this technical point~\cite{onion}.

A good way of thinking of the cluster detection is the
following~\cite{white03}.
First you use all the very high and very low
frequencies to remove foreground contamination. Next take the {\em
difference} between the 353 GHz and the 217 GHz channels. These two
channels have the same angular resolution, and since the SZ effect
roughly vanishes at 217 GHZ, then the resulting map will be a fair
representation of the clusters. That's it! We now have a full sky
map of all the clusters, which are simply sitting at the maxima of this 
difference-map. This method is really simplistic, and this is just
a good way of thinking of the procedure.

Next you naturally have to start asking how well this procedure
works. What is the {\em completeness}, that is how many of all the
clusters did you actually find? Second, what is the {\em reliability},
that is how many non-existing clusters did you find? With these two
numbers you get a good measure of your cluster detection algorithm,
and one can start discussing which is the optimal. Today very advanced
methods have been developed, see e.g.
\cite{schulz,geisbuesch,schafer1,schafer2,moscardini}.

We should dwell a while on the question of completeness and
reliability.  In order to get these two important numbers we need a
full {\em test} sky map of clusters. Basically we need to know what
Planck realistically will be seeing, and since a numerical simulation
of a full universe with both dark matter and baryons is impossible
today (by far!), then we have to do something smarter.

First we have to get realistic clusters and peculiar velocities, and
to that end we must make a full sky simulation of the dark matter
structures~\cite{kay01}.  Collisionless dark matter is relatively simple
to simulate, so this will allow up to $10^9$ particles in the simulation.
This will provide the basic structure of the entire universe, namely
a description of the positions and peculiar velocities of all the
gravitationally dominating dark matter.

Next you need to include the baryons. After all, the SZ effect is a
measure of the distribution of the baryons and electrons, so this
should be done as carefully as possible. For a given total mass of the
dark matter structure we know fairly well the radial distribution of
the dark matter~\cite{nfw,moore,jeans}. Thus, you just need for 'each'
of the DM structures in the full universe simulation to make another
simulation including the baryons. This is naturally impossible, so
instead one makes the following smart trick. We simulate {\em several}
clusters of different masses.  In reality we only simulate relatively
few, but we do those very carefully.  Then for a given cluster from
the full DM simulation we can 'interpolate' between these accurately
calculated clusters. This procedure is astounding, in my view, and the
results are impressive! The most recent
implementations~\cite{schafer1,geisbuesch} thus provide a very
realistic full SZ sky, both thermal and kinetic.

Let us get the perspective back. We now have a realistic full SZ 
sky, and we can use this to develop different methods of finding
clusters. Even better, we can compare the methods and find which one
is the best for Planck. The conclusion is roughly that Planck 
will detect about $10^4$ clusters out to redshift slightly beyond $z=1$,
with fairly good completeness and reliability.

Thus Planck will create a beautiful catalogue of clusters, which
basically includes all the massive clusters all the way out to
redshift of 1 or $1.5$. This catalogue is larger, deeper and has more
sky-coverage than any other existing catalogue. This cluster catalogue
will be very useful for future follow-up observations. One can
e.g. imagine a targeted SZ study, which selects the 100 brightest clusters
in the universe
for a careful study of radial temperature and density profile, which
then can be compared with numerical simulations. We remember that
the SZ observations are particularly powerful for large radii and for
distant clusters. The last is because the SZ is a distortion of the
CMB and hence basically redshift independent, and the 'large radius' is
because the SZ effect is proportional to the electron density, $I_{SZ}
\sim n_e$, whereas the X-ray emission is proportional to the electron
density squared, $S_{X} \sim n_e^2$.  I wish to repeat and emphasize
that {\em this catalogue in itself is very useful}, and we should do
our best to make it as accurate as possible!

Now, having established that the catalogue will come into existence,
let us proceed and discuss how accurately we will measure the 
cluster parameters of the {\em individual clusters}.

\section{Cosmology from the cluster catalogue?}
Basically, when we know the number, $n(M)$, of clusters of mass M, as
a function of redshift z, then we know the cosmology. Thus if we have
a complete and reliable catalogue which gives us $n(M)$, then we can
extract cosmological parameters like the matter density, $\Omega_M$,
cosmological constant, $\Omega_\Lambda$ etc etc (see
e.g.~\cite{carlstrom} for references).  I personally believe that most
of the studies conducted so far (which I do not cite here) are very
optimistic, and I do not advise people to blindly perform parameter
extraction with the cluster catalogue. Some of the reasons are the
following.

\begin{figure}
\begin{center}
\epsfxsize=8.5cm
\epsfysize=6.5cm
\epsffile{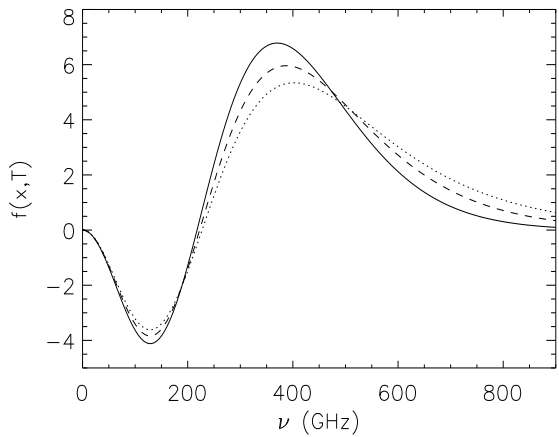}
\end{center}
\caption{The dilution effect from the relativistic corrections. 
The solid line is the non-relativistic effect (cold cluster), and the
dashed (dotted) line is for a 10 keV (20 keV) cluster.
Figure taken from ref.~\cite{diego}.}
\label{fig:dilute}
\end{figure}

The first point is naturally that one must understand the completeness
and reliability of the catalogue very accurately, in order to get
realistic error-bars. Many people have studied these numbers, however,
there are always effects which are not included, and the {\em
systematic} error-bars from non-included effects are very hard to
estimate.  For instance, the relativistic corrections to the SZ effect
will {\em dilute} the SZ signal at 353 GHz, which implies that a given
cluster will appear {\em less} bright than expected (see
fig.~\ref{fig:dilute}).  This will be a problem for the marginally
detectable clusters; we may think we should see them, but because of
the dilution effect they may be just non-detectable. Even worse, the
dilution effect is largest for the large and bright clusters, which
are very important for cosmological parameter extraction.

Another point is the question of scaling-relations. Basically we have
to make an assumption about the connection between temperature and
mass, the famous M-T relation. But what do we use here? Do we use the
observed M-T relation from X-ray observations? No, because the weight
from non-isothermal clusters is different between X-ray observations
and SZ observations. Do we use a theoretical M-T relation or one from
simulations? I think this would be very dangerous, after all these
relations are still not in very good agreement with observations, even
after years of effort.  And finally, in most of these studies one has
to make an assumption about the radial dependence of the baryons in
the clusters, and what do we choose - A beta-model for the 
baryons~\cite{cavaliere76}, or
a NFW profile~\cite{nfw}, or something completely 
different~\cite{hansenstadel,blois}? Basically my worry
is, that it is unclear today how to realistically estimate the {\em
systematic} error-bars from an insufficient (or incorrect) assumed
profile.

Thus, I personally don't believe we are in a position to use the
Planck cluster catalogue for a measurement of cosmological parameters
yet - there is still much work needed before we reach that level.  I
should clarify that all those objections above are my personal view,
and that many scientists, even very good ones, believe that I am being
overly pessimistic. However, this leaves space for pleasant surprises
down the data-analysis road.

\section{Cluster parameters from single clusters}
Having read the section above you may be asking yourself: Why don't
we simply measure the cluster temperature of the individual clusters
directly with the SZ? After all, the SZ effect depends on 3 parameter, 
the dominating Compton parameter $y$, the peculiar velocity, $v_p$,
and the cluster temperature, $T_e$. If we have multi-frequency observation
of the SZ effect, then we can separate the components and directly
measure the cluster temperature.

\begin{figure}
\begin{center}
\epsfxsize=8.5cm
\epsfysize=6.5cm
\epsffile{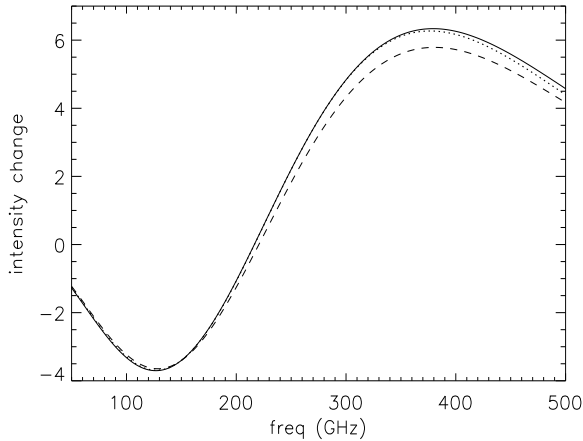}
\end{center}
\caption{Two clusters with different temperatures and peculiar velocity,
$T =5$ keV, $v_p=-145$ km/sec
and $T=7$ keV, $v_p=-260$ km/sec (full and dotted lines) are virtually
indistinguishable, whereas the cluster with 
$T=6$ keV, $v_p=0$ km/sec (dashed line) is clearly different.
Figure taken from ref.~\cite{aghanim03}.}
\label{fig:deg}
\end{figure}

The first problem is that some of the cluster parameters are degenerate.
This means that two clusters with {\em different} temperature, peculiar
velocity and Compton parameter, may produce virtually {\em indistinguishable
spectra}. An example hereof is given in fig.~\ref{fig:deg} where two
clusters (full and dotted lines) have parameters, $T =5$ keV, $v_p=-145$ km/sec
and $T=7$ keV, $v_p=-260$ km/sec respectively. These two lines are
essentially on top of each other. In comparison, the cluster (dashed line)
with $T=6$ keV, $v_p=0$ km/sec is clearly different.

Another problem is that the SZ signal is contaminated. The signal
which we have to analyse is a sum of different components. First
there is the SZ signal, but there is also interstellar dust emission,
infra-red galaxies and radio sources.  Naturally there is also a 
contamination from the CMB itself. All these contaminations have to
be either monitored and removed, or controlled and extracted
\cite{tegmark,aghanim97,vielva,herranz02,stolyarov,holder02,whitemajum,knox,nabilaguilaine}.

At the end of the day you will {\em still} be left with a little bit
of contamination - it is simply not possible to remove everything.
A specific example could be that you {\em assumed} that the non-removed
radio sources (or the dusty galaxies) would have a specific spectral
behaviour - e.g. a simple power-law. This may be a {\em wrong 
assumption} and you would systematically be biasing your results.

\begin{figure}
\begin{center}
\epsfxsize=8.5cm
\epsfysize=6.5cm
\epsffile{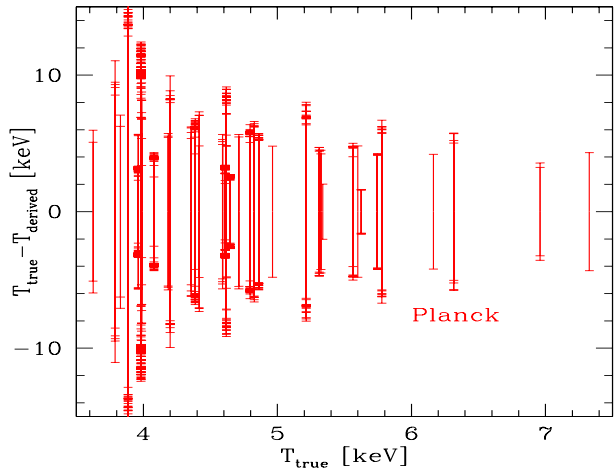}
\end{center}
\caption{The temperature of individual clusters will have larger
error-bars.}
\label{fig:Pt}
\end{figure}

How would this affect the SZ parameters? Let's assume that you have a
little bit of non-removed radio sources. That means that you have
removed most of the radio sources, but there is still a little bit you
didn't completely remove, maybe because you made the simplifying
assumption that it follows a power-law.  A contamination from radio
sources gives a positive contribution at low frequencies, where the
dominating $y$ parameter is determined, leading to a smaller value for
$y$.  Now, with this smaller value for $y$, the high frequency signal
seems too high, which can only be compensated with a very small
temperature.  Finally, the velocity term goes like $v_p/T_e$, so with
a very small temperature, $v_p$ is also forced to be very small. In
reality, this leads approximately to $T_e \approx v_p \approx 0$.

Similarly, the effect on the SZ parameters from 
a non-removable IR galaxies or dust emission is easy to understand,
see ref.~\cite{nabilaguilaine}.

Looking at fig.~\ref{fig:Pt} we see the final result for Planck
cluster temperature determination.   The
conclusion is that Planck will not be able to extract the temperature
for individual clusters, except for the hottest ones.  It is worth
remembering that {\em if} one has the courage to make specific
assumptions for the spectral behaviour of non-removed dust or point
sources, then one can decrease the error-bars on fig.~\ref{fig:Pt},
and hence extract the cluster temperature for individual clusters.

\begin{figure}
\begin{center}
\epsfxsize=8.5cm
\epsfysize=6.5cm
\epsffile{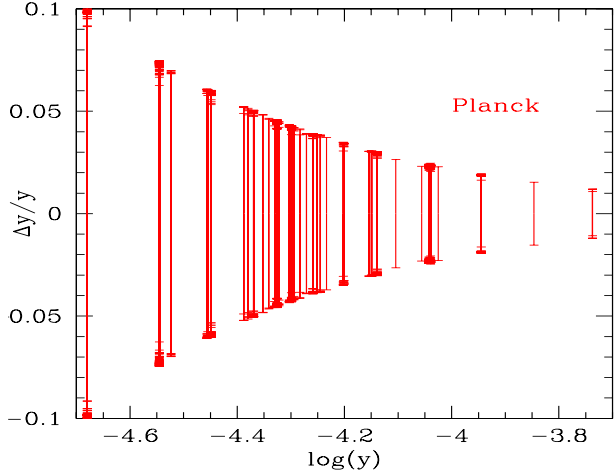}
\end{center}
\caption{The Compton parameter, $y$, as Planck will measure for individual
clusters. The brightest clusters will be measured to about $1\%$, 
and the dimmer clusters to about $10\%$.}
\label{fig:Py}
\end{figure}

The discussion above concludes that we cannot expect to measure the
sub-dominant SZ parameters, however, what about the dominating
Compton parameter, $y$. Fortunately this {\em can} be measured 
by Planck, and even to a very good precision.

The Compton parameter will be measured to about $1\%$ for the 
brightest clusters, and to about $10\%$ for the dimmer clusters
(see fig.~\ref{fig:Py}).
This is very good news, because these numbers were obtained under
the (rather pessimistic) assumption that we don't know the spectral
behaviour of all the non-removed contaminations. Thus, the accuracy
with which Planck will measure $y$ may be even better.

We therefore conclude that Planck will measure a very large number of
clusters, of the order 10 thousand, and the Compton parameter for each
individual clusters will be measured to a few percent. This is what
will give us a great catalogue for future targeted SZ experiments.
All in all, Planck will measure the CMB accurately, and also provide a
cluster catalogue which is larger, deeper and has more sky-coverage
than any other existing catalogue.  Naturally, one can use Planck
for investigating less conventional (more funny?) physics, such as
asteroids~\cite{Cremonese}, highly relativistic
electrons~\cite{ensslin} etc, but I will leave that discussion for
another good day.

\section{Conclusions}
The Planck satellite is a true piece of art, which will provide
impressive information about the CMB sky. 
The SZ signal can easily be removed, and will not be a problem
for CMB parameter extraction.

Planck will also give us
a cluster catalogue which is larger, deeper and has more
sky-coverage than any other existing catalogue. The Compton 
parameter for the individual clusters in this catalogue will be
measured to $1\%$ accuracy for the brightest clusters, and to
about $10\%$ for the dimmer clusters. Such a catalogue is very
useful for future targeted SZ studies.

I have argued why I personally believe that the systematic errors for
the individual clusters are still so hard to assess, that cosmological
parameter extraction with this cluster sample will be very difficult.
My view is rather conservative, so this leaves space for pleasant
surprised when we will analyse the coming Planck data carefully.

\acknowledgments
It is a pleasure to thank the organizers of the 2004 Varenna school
for excellent organization, the Planck collaboration (and in
particular Nabila Aghanim) for trusting me to give this lecture, and
the Tomalla foundation for financing my research in Switzerland.


\begin{thebibliography}{0}

\bibitem{planck} \BY{The Planck collaboration}
{\tt http://astro.estec.esa.nl/Planck}

\bibitem{balbi}  \BY{Balbi~A. et al.}
  \IN{ApJ}{588}{2003}{L5}

\bibitem{melch} \BY{Melchiorri~A.}
arXiv:hep-ph/0311319.

\bibitem{bowen} \BY{Bowen~R. et al.}
  \IN{MNRAS}{334}{2002}{760}


\bibitem{lfi} \BY{Mennella~A. et al.}
  \IN{AIP Conf.\ Proc.}{703}{2004}{401}

\bibitem{hfi} \BY{Lamarre~J.~M. et al.}
\IN{New Astronomy Reviews}{47}{2003}{1017}

\bibitem{villa}  \BY{Villa F. et al.}
\IN{AIP conf. Proc.}{616}{2002}{224}


\bibitem{Yurchenko}  \BY{Yurchenko V., Murphy J.~A. and Lamarre J.~M.}
arXiv:astro-ph/0205269

\bibitem{naselsky}  \BY{Chiang L.~Y. and Naselsky P.}
arXiv:astro-ph/0303140



\bibitem{aghanim97}  \BY{Aghanim N. et al}
\IN{A\&A}{325}{1997}{9}

\bibitem{white03}   \BY{White M. J.}
  \IN{ApJ}{597}{2003}{650}

\bibitem{geisbuesch} \BY{Geisbuesch~J., Kneissl~R. and Hobson~M.}
arXiv:astro-ph/0406190.


\bibitem{rephaeli01} \BY{Rephaeli~Y.}
arXiv:astro-ph/0110510.


\bibitem{sadehrephaeli} \BY{Sadeh S.  and Rephaeli~Y.}
  \IN{New Astron.}{9}{2004}{159}


\bibitem{hannestad}  \BY{Hannestad~S. et al}
   \IN{Astropart.\ Phys.}{17}{2002}{375}

\bibitem{bouchet} \BY{Bouchet F. R. and Gispert R.}
\IN{New Astron.}{4}{1999}{443}


\bibitem{heranz02}  \BY{Herranz D. et al}
\IN{MNRAS}{336}{2002}{1057}


\bibitem{diego} \BY{Diego~J.~M., Hansen~S.~H. and Silk~J.}
   \IN{MNRAS}{338}{2003}{796}




\bibitem{heranz04}  \BY{Herranz D. et al}
arXiv:astro-ph/0406226.


\bibitem{aghanimforni}  \BY{Forni~O. and Aghanim~N.}
arXiv:astro-ph/0402333.



\bi{sz72}  \BY{Sunyaev~R.A. and Zel'dovich~Ya.B.} 
  \IN{Comments Astrophys. Space Phys.}{4}{1972}{173}

\bibitem{gull76}   \BY{Gull~S. F. and  Northover~K. J. E}
  \IN{Nature}{263}{1976}{572}

\bibitem{lake77}   \BY{Lake G. and  Partridge R. B.}
  \IN{Nature}{270}{1977}{502}

\bibitem{laroque02}   \BY{LaRoque S.J. et al}
arXiv:astro-ph/0204134

\bibitem{coma}   \BY{De Petris M. et al}
  \IN{ApJ}{574}{2002}{L119}

\bibitem{wright79}   \BY{Wright E. L.}
   \IN{ApJ}{232}{1979}{348}

\bibitem{dolgov00} \BY{Dolgov A. D. et al.}
  \IN{ApJ}{554}{2001}{74}


\bibitem{rephaeli95} \BY{Rephaeli Y.}
  \IN{ApJ}{445}{1995}{33}

\bibitem{shimon}   \BY{Shimon M. and Rephaeli Y}
  \IN{New Astron.}{9}{2004}{69}

\bibitem{ItohV} \BY{Itoh N. et al} 
  \IN{MNRAS}{327}{2001}{567}

\bibitem{nozawa}  \BY{Itoh N. and Nozawa S.}
\IN{A\&A}{417}{2004}{827}

\bibitem{pointec}  \BY{Pointecouteau E., Giard M.,  Barret D.}
   \IN{A\&A}{336}{1998}{44}

\bibitem{hansen02}  \BY{Hansen S. H., Pastor   S. and Semikoz  D. V.}
  \IN{ApJ}{573}{2002}{L69}

\bibitem{sasz}  \BY{Hansen S. H.}
  \IN{New Astronomy}{9}{2004}{279}


\bibitem{onion}  \BY{Hansen S. H.}  
  \IN{MNRAS}{351}{2004}{L5}



\bibitem{schulz}  \BY{Schulz A.~E. and White, M.~J.}
  \IN{ApJ}{586}{2003}{723}

\bibitem{schafer1}   \BY{Schafer B. M. et al}
arXiv:astro-ph/0407089



\bibitem{schafer2} \BY{Schafer B. M. et al}
arXiv:astro-ph/0407090

\bibitem{moscardini}  \BY{Moscardini L. et al.}
\IN{MNRAS}{335}{2002}{984}



\bibitem{kay01} \BY{Kay S.~T., Liddle  A.~R. and Thomas P. A.}
\IN{MNRAS}{325}{2001}{835}


\bibitem{nfw} \BY{Navarro J.~F., Frenk  C.~S. and White S.~D.~M.}
\IN{ApJ}{462}{1996}{563}

\bibitem{moore}   \BY{Moore B. et al}
\IN{ApJ}{499}{1998}{L5}


\bibitem{jeans} \BY{Hansen  S.~H.}
\IN{MNRAS}{352}{2004}{L41}


\bibitem{carlstrom}  \BY{Carlstrom J.~E., Holder G.~P. and Reese E.~D.}
\IN{Ann.\ Rev.\ Astron.\ Astrophys.}{40}{2002}{643}


\bibitem{cavaliere76}  \BY{Cavaliere and Fusco-Femiano}
\IN{A\&A}{49}{1976}{137}


\bibitem{hansenstadel}  \BY{Hansen S. H. and Stadel J.}
\IN{ApJ}{595}{2003}{L37}

\bibitem{blois}   \BY{Hansen S. H.}
arXiv:astro-ph/0310302



\bibitem{aghanim03}  \BY{Aghanim N. et al}
   \IN{JCAP}{0305}{2003}{007}



\bibitem{tegmark}  \BY{Haehnelt M.~G. and Tegmark M.}
\IN{MNRAS}{279}{1996}{545}

\bibitem{vielva} \BY{Vielva P. et al}
 \IN{MNRAS}{344}{2003}{89}

\bibitem{herranz02}   \BY{Herranz et al.}
D.~Herranz, J.~Gallegos, J.~L.~Sanz and E.~Martinez-Gonzalez,
\IN{MNRAS}{334}{2002}{533}

\bibitem{stolyarov}  \BY{Stolyarov  V. et al.}
\IN{MNRAS}{336}{2002}{97}

\bibitem{holder02} \BY{Holder G. P.}
\IN{ApJ}{580}{2002}{36}


\bibitem{whitemajum}  \BY{White M.~J. and Majumdar S.}
\IN{ApJ}{602}{2004}{565}

\bibitem{knox}   \BY{Knox L, Holder G.~P. and Church S.~E.}
\IN{ApJ}{612}{2004}{96}



\bibitem{nabilaguilaine}  \BY{Aghanim N., Hansen S. H. and Lagache  G.}
arXiv:astro-ph/0402571


\bibitem{Cremonese} \BY{Cremonese G. et al.}
 \IN{New Astron.}{7}{2002}{483}


\bibitem{ensslin}  \BY{Ensslin T. A. and Hansen S. H.} 
arXiv:astro-ph/0401337





\end{thebibliography}
\end{document}